\documentclass[flalign,12pt]{wlscirep}

\usepackage{amsmath,amstext}
\usepackage[T1]{fontenc}
\usepackage{newtxmath,newtxtext}

\usepackage[figure,figure*]{hypcap}
\usepackage{hyperref}
\usepackage{booktabs}

\usepackage{float}

\newcommand{\overbar}[1]{\mkern 1.5mu\overline{\mkern-1.5mu#1\mkern-1.5mu}\mkern 1.5mu}

\title{A Data Driven, Zero-Dimensional Time Delay Model with Radiative Forcing for Simulating Global Climate }

\author[1,2,*]{Rajashik Tarafder}
\author[1,2]{Dibyendu Nandy}
\affil[1]{Center of Excellence in Space Sciences India }
\affil[2]{Department of Physical Sciences, Indian Institute of Science Education and Research, Kolkata}
\affil[*]{rajashiktarafder@gmail.com}
\begin{abstract}
Several complicated non-linear models exist which simulate the physical processes leading to fluctuations in global climate. Some of these more advanced models use observations to constrain various parameters involved. However, they tend to be very computationally expensive. Also, the exact physical processes that affect the climate variations have not been completely comprehended. Therefore, to obtain an insight into global climate, we have developed a physically motivated reduced climate model. The model utilizes a novel mathematical formulation involving a non-linear delay differential equation to study temperature fluctuations when subjected to imposed radiative forcing. We have further incorporated simplified equations to test the effect of speculated mechanisms of climate forcing and evaluated the extent of their influence. The findings are significant in our efforts to predict climate change and help in policy framing necessary to tackle it.
\end{abstract}
\begin{document}

\flushbottom
\maketitle
\thispagestyle{empty}
Reconstructions of global mean temperature show a quasi-periodic change in climate throughout Earth's geological history. This change usually consists of a rise and fall in temperature owing to small differences in the amount of solar radiation that reaches the Earth. The effect has been attributed to the change in the eccentricity of Earth's orbit around the sun, change in tilt angle  and the precession of its tilt axis\cite{Hays1976}. This phenomenon is referred to as the Milankovitch Cycle.

However, the current period of warming remains of enormous significance and interest because of its unexpected rate of change. This sudden increase in temperatures across the planet has raised concerns that we may be at the peak of the Holocene extinction period (Sixth Great Mass Extinction). Studies have concluded that between 15 to 37\% of endemic plants may become extinct by 2050 \cite{thomas2004}. The impacts are not only significant but probably indicate towards irreversible changes in the biodiversity of the planet. This increased rate of temperature change is thought to be human induced \cite{Santer1996}. Such a conclusion is arrived at from consensus exhibited by a large number of climate models that show a significant agreement with observed changes in climate. The models point toward a greenhouse forcing modulated climate.  

Though enormous scientific consensus exists on the ability of the models to produce credible quantitative estimates of future climate change, there exist typical areas of low-confidence in certain finer aspects of the models and their scientific basis. It is recognized that the scientific understanding of processes such as cloud albedo effect, solar irradiance, volcanic aerosols, cosmic rays, etc. remain significantly low \cite{ipcc2007}. The presence of these uncertainties hinders our ability to successfully predict future climate change and mitigate our response towards the same. Therefore, it remains scientifically important that we study the impact of these various factors on our climate system.

The fluctuations in the earth's climate is a result of energy imbalance created by a variety of different processes. These processes can be broadly classified to be interacting with the atmosphere along one of the three pathways - absorption and release of incoming radiation by various parts of the earth,  the reflection of incoming shortwave radiation into space, and the absorption of outgoing longwave radiation. In the last century, we have studied and quantified the major processes that act on these pathways.

From 1960 to 1980, a dimming of global shortwave radiation was observed \cite{Wild2005}. This dimming was succeeded by a similar brightening in around 75\% of the observation stations, causing global warming \cite{Ohmura2006}. This second period coincided with a reduction of $SO_{2}$ emissions into the atmosphere from the industrial world by about 2.7\% per annum\cite{Stern2005}.

It was found that $SO_{2}$ is oxidized in the atmosphere to produce sulfates \cite{Hegg1967}; which belong to a class of atmospheric chemicals known as aerosols. Aerosols impact the atmosphere's radiation budget by directly scattering incoming solar radiation back into space and indirectly by acting as Cloud Condensation Nuclei (CCN) \cite{Haywood2000}. Of these, the presence of higher number of CCN leads to higher number of cloud droplets in a smaller region of the atmosphere. This increased droplet concentration further scatters more solar irradiance back into space. Further, larger Cloud Droplet Number Concentration (CDNC) leads to longer cloud lifetimes, as particles are not dense enough to produce precipitation \cite{Rosenfeld2000}. To fully appreciate the impact of aerosols, it is important to realize that the period of global cooling observed between 1940-1980 was due to higher radiative forcing by aerosols as compared to the greenhouse forcing during that time. Radiative forcing due to aerosols can range up to 2 $W/m^2$ (after adjusting for projection effects). The current  greenhouse forcing is $\sim 3W/m^2$ 

As opposed to the aerosol forcing which cools the earth's atmosphere, greenhouse forcing acts along the third pathway and increases its heat content\cite{houghton1992}. This process remains one of the most widely studied aspects of our climate system and has been attributed to causing the global warming effect. The increased forcing along this pathway is due to the increase in greenhouse gas concentrations in the troposphere. 

Though aerosols and greenhouse gases affect the total heat content of the atmosphere, the transient response of the climate to temporal variations in forcing is determined by the oceans\cite{randall2007}. This is because only a top few meters of ocean water holds as much heat as the entire atmosphere\cite{Levitus2009} and the response to incident forcing is slight. This becomes even more important as ocean circulation belts drag the heat into the depths of the ocean only to be reintroduced later\cite{wunsch2004}. This process introduces a delay in the earth's response to radiative forcing and has been popularly referred to as climate inertia. The inertia of the climate is widely accounted for in Global Circulation Models (GCMs), and the ocean's thermal inertia is believed to delay global warming for timescales in the range of several decades \cite{Hansen2013}. GCMs simulate complex circulation processes in the oceans to account for this inertia. These models provided significant breakthroughs in our understanding of the climate in the last century. 

However, GCMs are computationally expensive and often tend to overestimate the response of the climate (111 out of 114 models) when subjected to radiative forcings\cite{flato2013}. This situation is further aggravated due to the presence of several non-linear interactions that make the study of individual pathways a formidable task. To overcome the computational expense of GCMs when dealing with fluctuations in the global climate and to aid the investigations of climate through GCMs we introduce a zero-dimensional time delay model. Zero-dimension refers to the spatial independence we impose in our model i.e. we study the variation in globally averaged temperature vs. time.

However, the radiative forcings involved in our climate system are spatially heterogeneous. We mimic spatial transport and mixing in such a model by introducing time delays in the source terms of the energy balance equation for the atmosphere. This converts the ODE to a delay differential equation (DDE)\cite{driver2012}. The time delay involved is motivated from the period of circulation belts (these belts carry the heat into the depths of the oceans and reintroduce them after a finite delay). But due to the presence of different circulations with timescales ranging from a few weeks to several decades, the delay is expected to be a weighted contribution of various circulation systems. 

The notion of DDEs has been studied earlier in the context of Babcock-Leighton mechanisms in dynamos \cite{Yoshimura1978}$~$\cite{Wilmot2006}. Our model builds on the ideas developed in simulating dynamos with DDEs and extends it to climate systems.

\section*{Results}

To obtain the best fit condition, the delay term we introduced in our model is varied in between no delay to a maximum of two solar cycles. For each such delay, we perform a correlation analysis. The peak of the correlation vs. delay curve denotes the best fit of our model with the observed data. The corresponding delay thus represents the effective time delay due to heat circulating processes in the oceans. 

\begin{figure}
\captionsetup{singlelinecheck = false, justification=justified}
\includegraphics[width=\linewidth]{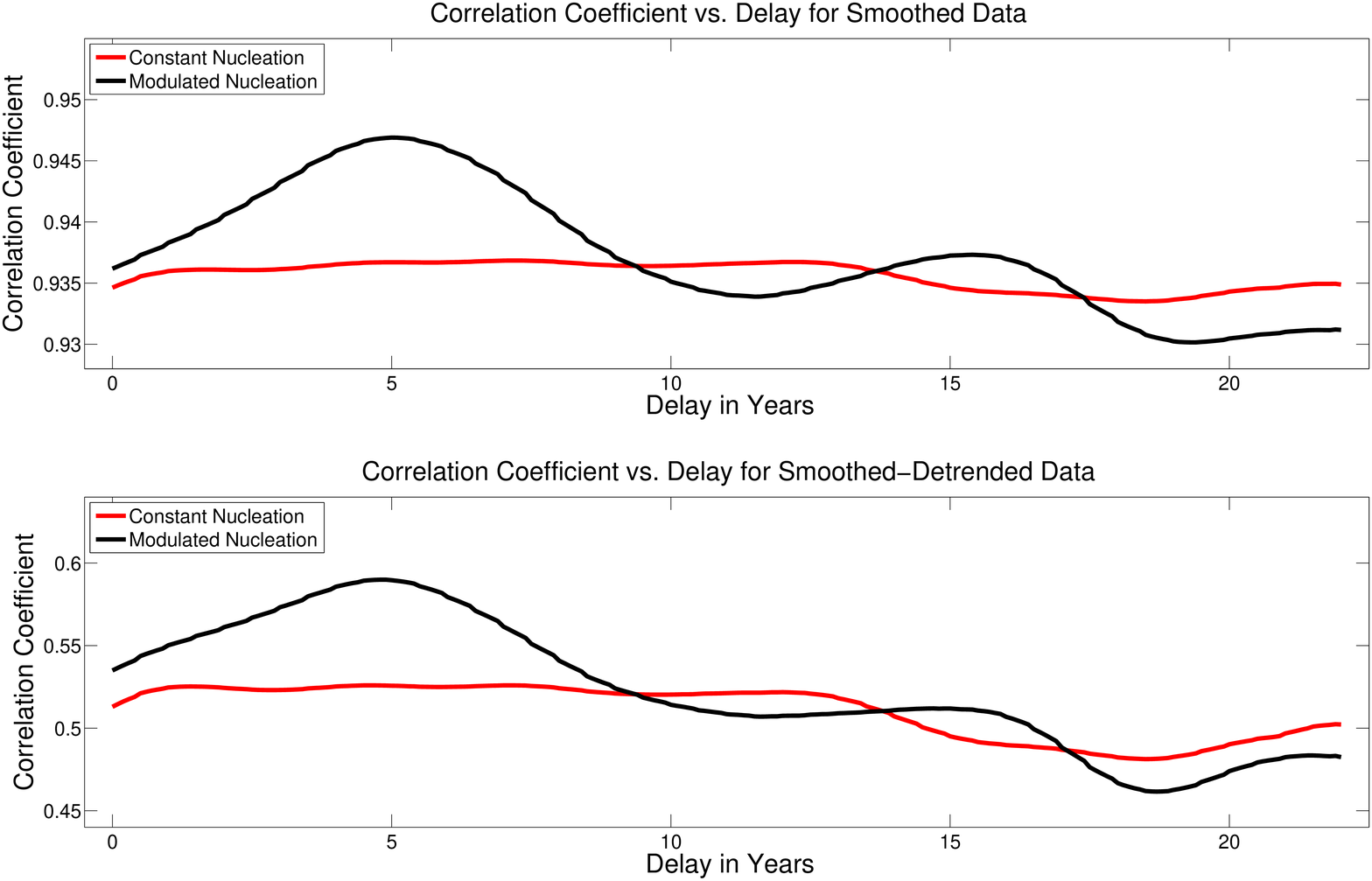}
\caption{The panels present the change in correlation obtained by varying the delay in our DDE. For the top panel, we have used smoothed data for both constant nucleation and modulated nucleation scenarios. The bottom panel uses the smoothed data which have also been detrended. The detrending has been conducted to amplify the difference in correlation value between constant nucleation and modulated nucleation. This is because the difference in the two scenarios for the top panel cannot be considered significant.}
\label{fig:corr}
\end{figure}

However, we observe that there is an inappreciable change in correlation when the model considers the solar forcing to be only due to the change in Total Solar Irradiance (TSI). On introducing the secondary effect, which modulates the aerosol nucleation rate, we notice that there is a distinguishable change in correlation with varying time delay. This procedure also improves the correlation obtained between the simulated and observed temperature. The above results have been demonstrated in the top panel of fig. \ref{fig:corr}.

\begin{figure}
\captionsetup{singlelinecheck = false, justification=justified}
\includegraphics[width=\linewidth]{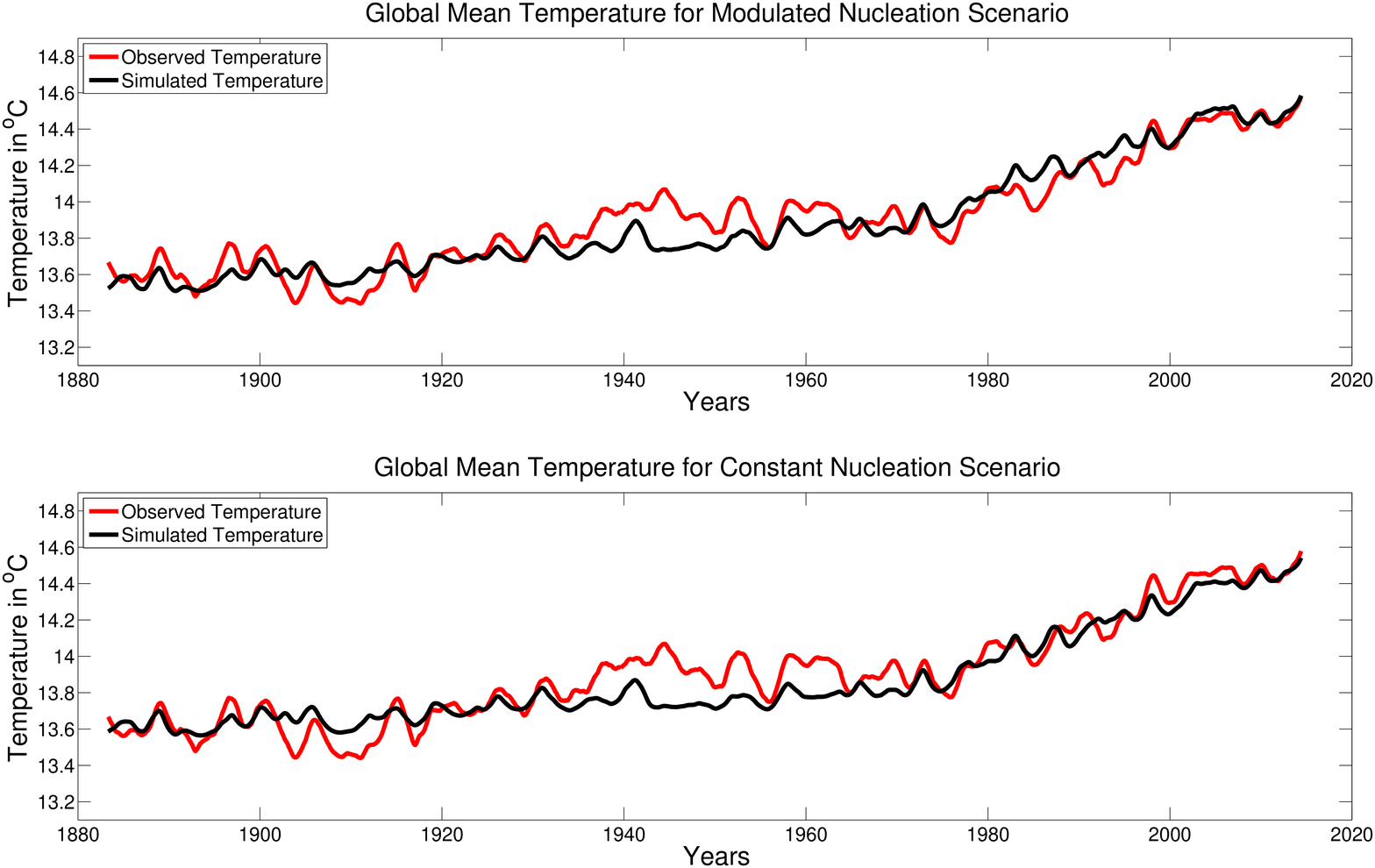}
\caption{The above panels demonstrate the best fit condition for the simulated and observed values of temperature. We have demonstrated the plots obtained for both modulated and uniform nucleation. The delay used for the top panel is 4.9 years whereas the bottom panel uses a 3.0 year delay. The bottom panel exhibits no significant difference for a delay of 0-11 years.}
\label{fig:temp}
\end{figure}

The maximum correlation thus obtained after the above operations is for a 4.9 year delay considering secondary modulation by solar activity. This best fit condition has been plotted in fig. \ref{fig:temp}.

t must be noted that the change in the value of correlation is nugatory for the two separate nucleation scenarios. This is because the largest contributor to the correlation value is the global trend. As the greenhouse gas forcing dominates this trend, the remaining terms only contribute to the fluctuations along the curve. Therefore, to distinguish the difference in the two scenarios, we detrend the data by removing the long-term trend due to greenhouse gas forcing. This has been demonstrated in the bottom panel of fig \ref{fig:corr}. The difference in correlation values is relatively higher (0.07) to consider the modulated nucleation as a finite contributor to climate fluctuations.    

The above detrending has been conducted using
\begin{equation}\label{eq:detrend}
Trend(t)=T_{0} +~Temperature~Change~Due~to~Greenhouse~Forcing 
\end{equation}
where
$T_{0}$ is the initial value of the simulated temperature time series.  

\begin{figure}
\captionsetup{singlelinecheck = false, justification=justified}
\includegraphics[width=\linewidth]{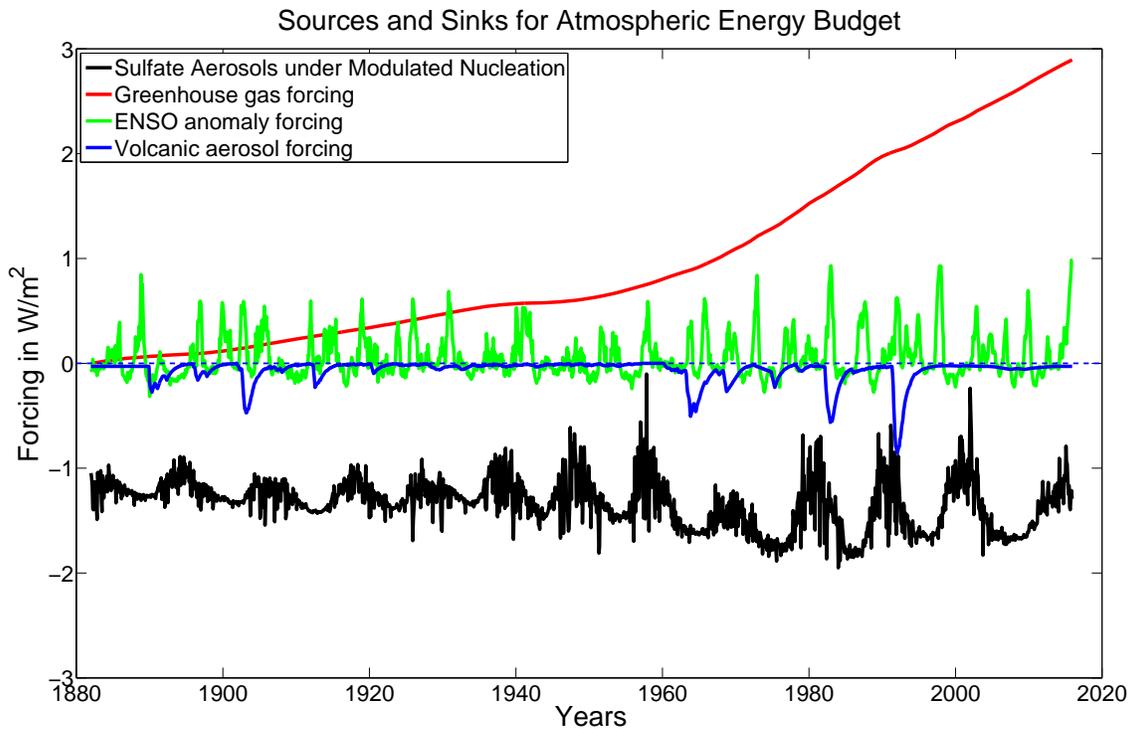}
\caption{The figure represents the global impact of the varied forcing terms. The impact due to ENSO and aerosol forcing have been scaled to account for the fractional area of the earth they contribute in. For example, albedo change due to aerosols only contributes to the area of the earth facing the sun (projected area is one-fourth of the total area). And ENSOs operate in the tropical region of the Pacific Ocean.}
\label{fig:forcing}
\end{figure}

The forcing terms involved have been plotted in fig. \ref{fig:forcing}. These terms are unaffected by the delay. 

\begin{figure}
\captionsetup{singlelinecheck = false, justification=justified}
\includegraphics[width=\linewidth]{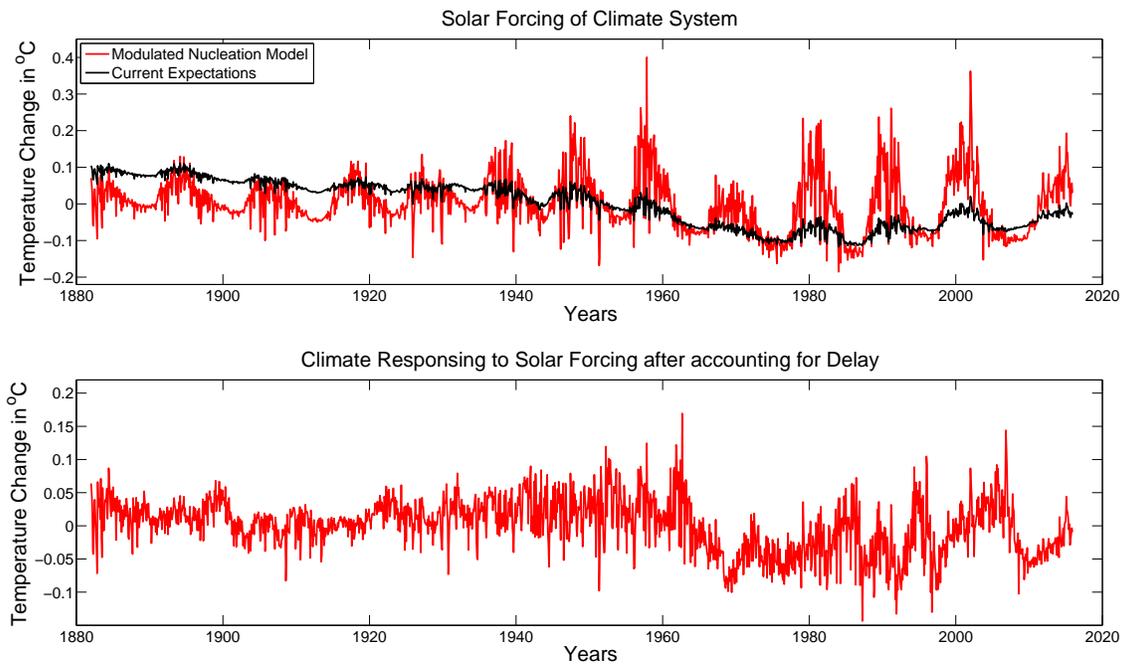}
\caption{The top panel demonstrates the temperature change of the atmosphere in response to solar forcing. The black curve shows our current estimate of solar forcing where the only impact is due to variance in TSI. The red curve denotes the case for modulated nucleation which enhances the impact of solar forcing on the atmosphere. However, due to the existence of delay in the recycling of heat by the oceans, this signature of an 11-year cycle is masked. The bottom panel shows the effective response after delay is included. Both the curves have been centered at zero by subtracting their mean value.}
\label{fig:solar}
\end{figure}

Due to the presence of a nucleation modulation mechanism, there is a need to reassess the impact of solar activity on our climate. Earlier estimates of solar forcing concluded a change in 0.07K in global temperature due to change in TSI over a solar cycle \cite{gray2010}. Using our best fit condition, we find that solar activity causes a change of 0.27K for modulated nucleation as compared to 0.08K obtained with our model using existing consensus. This has been demonstrated in the top panel of fig. \ref{fig:solar}. However, the discernible signature of the solar cycle that we see in the forcing gets masked due to the delay effect of oceans. The temperature response, therefore, remains higher than our current literature value but lower than the calculated value of 0.27K. This effective response is $\sim$ 0.17K. Therefore, it appears that the solar forcing maybe 2 to 3.5 times stronger than current estimates. The values for temperature change are obtained by averaging the $3\sigma$ range for temperature for each cycle from 13 to 23.       

However, during a period of prolonged low activity by the sun as in the case for Maunder Minimum, this delayed signature would be absent. This would lead to drop in solar forcing further causing a decrease in global temperatures of 0.3 K or more.

\subsection*{Prediction}

Turbulent pumping of magnetic flux in the solar interior restricts the predictability of solar cycles to only the next cycle \cite{karak2012}. We have therefore used recent estimates of the strength of cycle 25 to expand the TSI time series upto 2031. The strength of cycle 25 is expected to be similar to cycle 16 \cite{bhowmik2017}. We have extended the TSI time series using the TSI data for cycle 16.    

For the greenhouse gas levels, we have used the RCP 2.6 levels upto 2031. Further, tropospheric aerosols have been assumed to be constant at 2011 levels whereas stratospheric aerosols have been assumed to be non-existent. It must be noted that due the volcanic source of stratospheric aerosols, predictions for their levels are not achievable.  

\begin{figure}
\includegraphics[width=\linewidth]{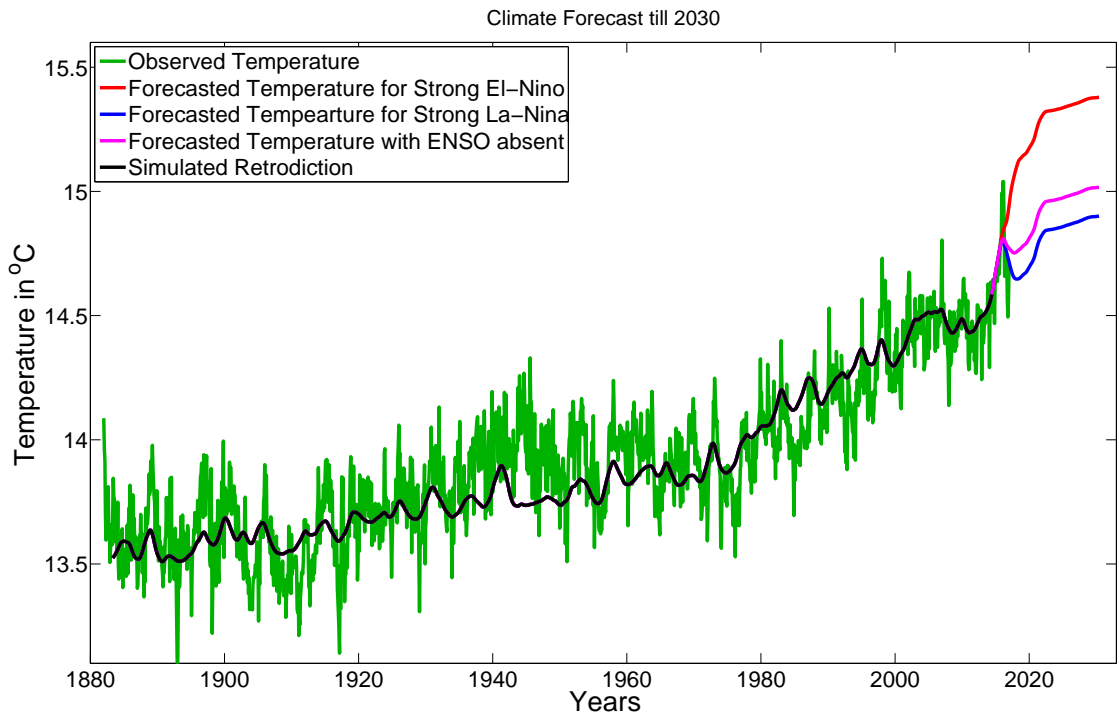}
\caption{The above plot demonstrates projected temperatures till 2030. The green curve denotes observed temperatures till 2016. The black curve denotes the simulated temperatures for which the best fit conditions were obtained. The remaining three curves are forecast scenarios - red represents the maximum temperature considering a perpetual strong El-Nino condition, blue represents the maximum temperature considering a perpetual strong La-Nina condition and the magenta curve demonstrates the case for absence of ENSO effect.}
\label{fig:predict}
\end{figure}

Similarly, our understanding of ENSO and its predictions remains rather uncertain. Most models are not able to predict the nature of ENSO beyond few months at a time. We have therefore calculated three different forecast scenarios for three different ENSO conditions. The forecasts obtained using the strong El-Nino conditions, and strong La-Nina condition represent the two extreme trends for change in temperature. The third scenario, in the absence of ENSO, represents the most probable trend for change in temperatures. All fluctuations are expected to be about this trend. Temperatures are expected to cross $15^0C$ according to this trend by 2030. These results have been demonstrated in fig. \ref{fig:predict}.

\section*{Discussion}

To summarize, we have constructed a physically motivated reduced climate model, which includes time delay (due to the circulation of heat by oceans), to study fluctuations in global climate. The model can be generalized to study a variety of forcing scenarios and effect of individual circulation belts by introducing an appropriate delay. The delay of 4.9 years obtained in our analysis is a weighted contribution from these different circulation belts. These belts have time periods ranging from a few months in the case of open ocean convection \cite{marshall1999} to several decades as in Pacific Decadal Oscillations \cite{mantua2002}.  

A significant result that has emerged from our model is the better correlations when modulation of nucleation rates by aerosols is considered. Although there has been much debate on the efficacy of such a modulation, support for a physical mechanism that causally connects this modulation to albedo has also started developing. The mechanism speculates that nucleation rates of water molecules in the atmosphere due to aerosols are strongly impacted by the strength of Galactic Cosmic Ray (GCR) flux \cite{Kirkby2011}. The nucleation rates are expected to increase up to ten folds or more (implying higher cloud cover) in the presence of GCR as compared to neutral radiation. Though this mechanism remains under considerable debate, authors have shown a correlation between increases in cosmic ray flux to the increase in cloud cover \cite{Svensmark1997}. This, however, was met with criticism \cite{Kuang1998} as similar high correlations were obtained on using El Nino Southern Oscillation. Further objections relating the absence of cloud seeding at expected altitudes were also raised \cite{Kernthaler1999}$~$\cite{Jorgensen2000}$~$\cite{Gierens1999}. Our analysis suggests that such a modulation mechanism certainly explains the global trends in temperature better. 

It must be noted here that experiments in laboratory conditions have also demonstrated the ability of GCRs to form ice nuclei \cite{Tinsley1996}$~$\cite{Pruppacher1973}$~$\cite{Abbas1969}. The need to understand the impact of this mechanism becomes even more important due to its modulation by solar activity \cite{balasubrahmanyan1969}. Cosmic ray flux is expected to decrease with increase in solar activity and vice versa. This would imply a higher radiative forcing due to variations in solar activities than has been accounted for in current models.

Though speculations have been raised about climate change being dominantly modulated by solar variability or galactic cosmic rays, it can be clearly observed from fig. \ref{fig:forcing} that none of these factors are stronger than the anthropogenic greenhouse forcing present. The figure clearly demonstrates the dominant nature of greenhouse forcing. It is this element that lends the climate its overall trend.  Another important feature that can be observed from the figure is that GCRs acting on aerosols contribute a larger forcing than that of ENSO anomaly. The trend, however, becomes less visible in the final temperature response due to the delayed impact from the oceans. This makes the detection of signatures of GCRs in global temperatures harder to detect.

Finally, we have coupled the temperatures at the top of the atmosphere to that of the base. This allows us to impose flux conservation on outgoing radiation. An immediate result of such a method is the increased radiative cooling of earth with increasing temperatures. This means the earth's efficiency to cool itself would gradually improve with increasing temperatures. However, such a process cannot overcome the greenhouse warming effect.

\section*{Methods}

\subsection*{The Model}\label{ssec:model}

The model tries to simulate fluctuations in globally averaged temperature by calculating the energy imbalance in the atmosphere. This imbalance is a result of various physical processes as discussed earlier. We attempt to capture the physics of these fluctuations by developing a DDE. The sources and sinks of energy are placed on the right-hand side while the left-hand side of the DDE denotes its corresponding response. The energy equation thus developed encodes the atmosphere's interactions with the earth's surface, the sun, and space. 

As a zero dimensional model requires spatial independence, we have made a few first order approximations. These include considering that the energy imbalance changes the temperature of the lower atmosphere uniformly. Further, as the lower atmosphere retains most of the heat of the atmosphere, we only consider temperature changes in this region in our calculations. These approximations jointly allow us to use the change in temperature at the base of the atmosphere as a single representative temperature in our equations.

The reduced delay differential equation is thus defined as
\begin{equation}\label{eq:main}
\begin{aligned}
C\frac{dT_{b}(t)}{dt}=&~a*\frac{A}{4}*I_{net}(t)+b*\frac{A}{4}*I_{net}(t-T_{0})+c*\frac{A}{4}*\epsilon*I_{net}(t-T_{0})\\
                      &-\sigma*A*(T_{TOA}(t))^{4}+G(t)*A+E(t)
\end{aligned}
\end{equation}
and
\begin{equation}\label{eq:Inet}
I_{net}(t)= I(t)(1-\alpha(t))
\end{equation}
where 
\begin{itemize}
\item I(t) refers  to the total solar irradiance at the top of the atmosphere.
\item $\alpha$(t) is used to describe global mean albedo. The variability in the value of albedo due to aerosols has been discussed later in the article.
\item C represents the heat capacity of the atmosphere. The value of this constant can be approximated to be the product of specific heat capacity of air at 287K \cite{Hilsenralh1956} and mass of the atmosphere \cite{Trenberth2004}. The value of C thus obtained is $\sim 5.1*10^{21}J/K$.
\end{itemize} 

The first source term on the right-hand side of the DDE represents the fraction of Total Solar Irradiance (TSI) directly absorbed by the atmosphere. As this process occurs at the time of passage of solar radiation through the atmosphere, no delay is expected in its corresponding influx of energy.

The earth's surface then absorbs the remainder of the TSI (after absorption in the atmosphere). As the continents do not circulate heat and only trap the heat in the upper layers of the soil, they are almost immediately released back into the atmosphere. However, the oceans have far higher heat capacity and store an equivalent amount of the atmosphere's heat in its top few meters \cite{Levitus2009}. In addition to this, the strong circulation belts in the ocean drag the heated water into deeper layers \cite{wunsch2004}. Further, as oceans and other water bodies cover a major fraction of the earth's surface, the effective response of the surface is dominated by that of the oceans. This causes the heat that reaches the surface to be released after a finite delay.The second and third terms of the DDE encode this effectively delayed response as explained below. 

The release of heat from the surface occurs by two physical processes:
\begin{itemize}
\item The evaporation of water from the ocean surfaces transfers energy to the atmosphere in the form of latent heat. This heat is gained by the atmosphere during the condensation of clouds and is denoted by the second term.
\item The surface of the Earth releases heat in the form of blackbody radiation. This released energy is trapped by the atmosphere due to the greenhouse effect. As the forcing due to increase in greenhouse gasses (industrial period) is separately included, we consider the pre-industrial forcing alone while expressing the third term.
\end{itemize}	

The radiative flux of the Earth's surface is accompanied by a back-radiation from the atmosphere. The net radiative flux entering the atmosphere, as a result of these two processes, is expressed by the third term.    

The values of a, b, c and $\epsilon$ are obtained using the energy budget at the surface of the earth \cite{Trenberth2008}  and at the top of the atmosphere \cite{Loeb2009}. The values and their physical interpretation have been listed in Table. \ref{tab:model_param}. 

\begin{table}[h!]
\centering
\caption{Parameter values for zero-dimensional model}
\label{tab:model_param}
\begin{tabular}{lllll}
 \toprule
 Constant & Value & Physical Interpretation\\
 \midrule
 a  & 0.33 & fraction of Total Solar Irradiance (TSI) \\
 &          & directly absorbed by atmosphere.\\
 b & 0.40 & fraction of TSI reintroduced into \\
 &          & atmosphere through evaporation, \\
 &          & conduction and convection.\\
 c & 0.27 & fraction of TSI radiated back to the \\
 &          & atmosphere by earth. This is the net \\
 &          & fraction after emission from surface\\
 &          & and back radiation.\\
 $\epsilon$ & 0.33 & effective pre-industrial greenhouse gas\\
 & & capture of longwave radiation.\\
 \bottomrule
\end{tabular}
\end{table}

The fourth term in the DDE denotes the loss of heat by the atmosphere due to blackbody radiation. As the atmosphere is optically thick towards infrared radiation \cite{sloan1955}, the radiative loss is primarily due to the outer layers of the atmosphere. We, therefore, require an expression for the temperature at the top of the atmosphere and relate it to the temperature at its base.

The expression for temperature the top of the atmosphere $(T_{TOA})$ is obtained by solving the radiative transfer equation using the Grey atmosphere assumption along with the Eddington approximation. This approximation can be interpreted as blackbody radiation occurring from the skin temperature of the approximately isothermal outer boundary of the atmosphere \cite{Sagan1972}. This results in
\begin{equation}
T_{TOA}(t)=\frac{T_{b}(t)}{2^{1/4}}
\end{equation}
where $T_{b}$ is the temperature at the base of the atmosphere.

The variables in our model, therefore, are I(t), G(t), E(t) and  $\alpha$(t). A description of these quantities and the processes that affect them have been discussed in the following chapter.

\begin{figure}
\captionsetup{singlelinecheck = false, justification=justified}
\includegraphics[width=\linewidth]{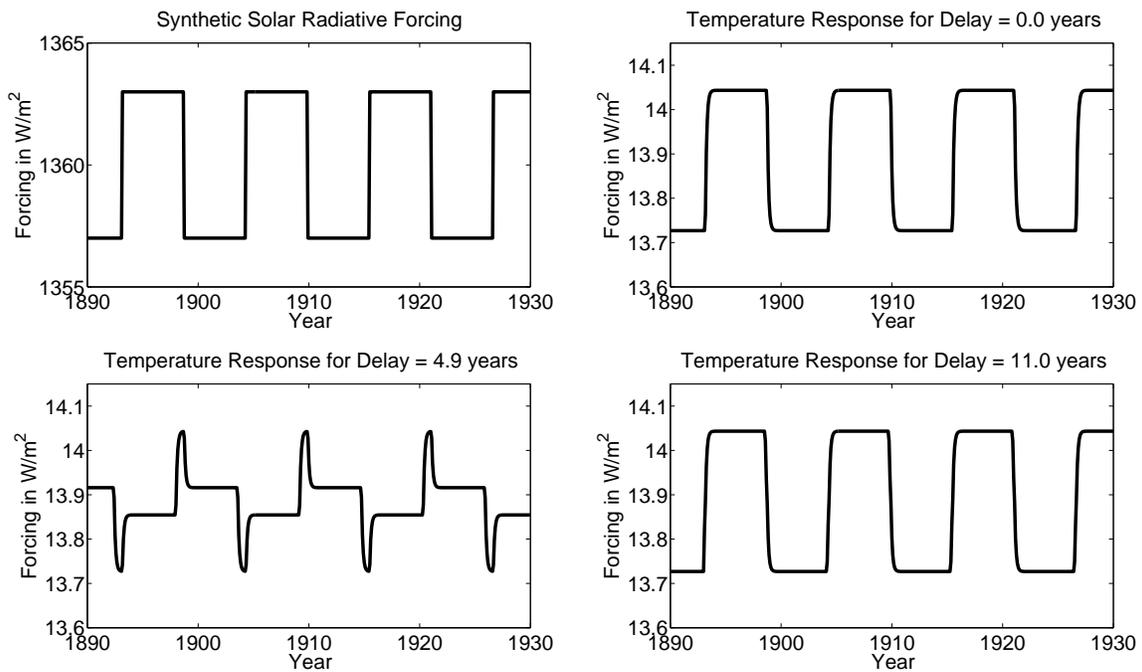}
\caption{The above panels demonstrate the effect of time delay on an imposed synthetic solar radiative forcing. The forcing has an 11 year period, and we see a response identical to the forcing in the case of zero delay and eleven year delay. The temperature profile for the delay of 4.9 years presents a more complicated response. This is because a fraction of the input forcing undergoes the effect of delay while the rest is instantaneously introduced into the system.}
\label{fig:dummy}
\end{figure}

Finally, the effect of the delay term introduced in our model has been demonstrated in Fig. \ref{fig:dummy} using synthetic solar irradiance forcing. 

\subsection*{Variables and Data}\label{ssec:data}
\subsubsection*{Aerosols}\label{sssec:aerosols}

In this model, we have used sulfate as a proxy for all anthropogenic aerosols in the atmosphere. This is because sulfate is the dominant source for CCN \cite{Charlson1992}. Also, sub-micrometer particles responsible for shortwave light scattering in the atmosphere is predominantly produced from the chemical reactions of sulfur compounds. In addition to this, the source of $SO_{2}$ emissions which in turn produce the sulfate aerosols is well constrained and relatively well known \cite{Stevens2014}.  

\begin{figure}
\captionsetup{singlelinecheck = false, justification=justified}
\includegraphics[width=\linewidth]{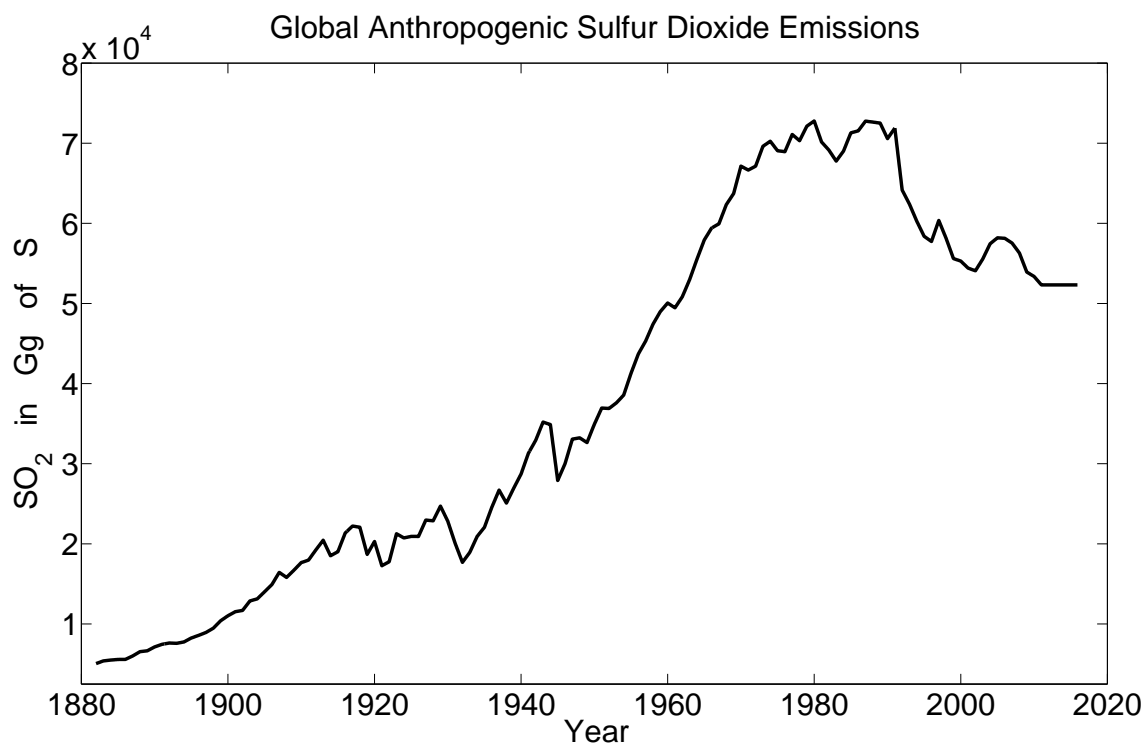}
\caption{The time series data for Sulfur Dioxide has been obtained from the calculations of D. Stern\cite{Stern2005} and, Z. Kilmot and his collaborators\cite{Kilmot2013}. Emission levels have been expressed in Giga-grams of sulfur and must be converted to equivalent $SO_2$ levels before being used in calculations. These values are calculated using the GDP values of all countries and the variety of industries that function within it. This is because the primary sources of anthropogenic sulfur dioxide emissions are industries and burning of fossil fuels.}
\label{fig:sulfur_emissions}
\end{figure}

In the absence of long-term data for sulfur dioxide levels in the atmosphere, we have used the calculated values of global aggregate sulfur dioxide emission for 1882-2000 \cite{Stern2005}. To extend the time series, we have used values calculated in other publications \cite{Kilmot2013}. Due to the differences in the emission levels calculated by the different authors, the values have been scaled using a constant factor to match that of the initial dataset for the year 2000. Emission levels beyond 2011 have been assumed to be steady at 2011 levels. As the time series is annual, a monthly data was obtained for our calculations by using a linear trend for growth/decline of emissions between January and December for any particular year. Global sulfur dioxide emissions after the above operations have been shown in Fig. \ref{fig:sulfur_emissions}. 

The use of emission data instead of sulfur dioxide levels poses no problems as residence timescale of sulfur compounds is on the scale of few days in the troposphere. Therefore, the emissions of one month have limited influence on the other.

Apart from the anthropogenic sources considered above, we have included sulfur burdening of troposphere due to natural sources. This is primarily contributed by Dimethyl Sulfide (DMS) emissions from oceans and $SO_2$ degassing at volcanic sites. While volcanoes contribute up to 9Tg (tera-grams) in S per year \cite{graf1997}, the DMS flux of oceans has been estimated to range between 13-37 Tg in S per year \cite{Kettle2000}. However, the efficiency of sulfate burdening due to these sources has been estimated to be 2.5 times and 4.7 times of the anthropogenic sources respectively \cite{graf1997}. We have, therefore, considered an equivalent background output of 80Tg of S in anthropogenic terms.

Using the obtained values of sulfur emission levels, we calculate the impact that aerosols have in the earth's energy budget. For this, we employ existing schemes \cite{Charlson1992} to measure direct and indirect effects of sulfate levels in the atmosphere. Further, latest estimates of effective cloud cover interacting with aerosols \cite{Stevens2014} have been used to revise the schemes and obtain the change in albedo. The calculations involved have been detailed below.\\ 
\\
\textit{For direct radiative forcing}
\begin{equation}\label{eq:sulf_dir_1}
\overbar{\Delta \alpha}=-k(1-A_{c})\overbar{\Delta R_{SO_{4}^{-2}}}
\end{equation}
where fractional cloud cover is denoted by $A_{c}$, and $R_{SO_{4}^{-2}}$ is the change in planetary mean albedo due to aerosol concentration. We have only considered the clear sky fraction for direct aerosol forcing as the optical sulfate burden is masked in areas of thick clouds and bright surfaces \cite{Stevens2014}.We further introduce a new term 'k' that models the fraction of sulfate molecules that effectively act as nuclei. For optical depth $\delta_{SO_{4}^{-2}} \ll 1$, we can represent $\overbar{\Delta R_{SO_{4}^{-2}}}$ as,
\begin{equation}\label{eq:sulf_dir_2}
\overbar{\Delta R_{SO_{4}^{-2}}} \cong 2T^{2}(1-R_{s})^{2}\overbar{\beta}\overbar{\delta_{SO_{4}^{-2}}}
\end{equation}
where T is the fraction of incident light that reaches the aerosol layer from the atmosphere above, $\beta$ is the fraction of irradiance scattered upwards by the aersol, $R_{s}$ is the mean albedo of the surface below the aerosol layer, and $\overbar{\delta_{a}}$ is the mean optical depth of aerosol. Combining equations \ref{eq:sulf_dir_1} and \ref{eq:sulf_dir_2} we obtain
\begin{equation}
\overbar{\Delta \alpha}=-2T^{2}k(1-A_{c})(1-R_{s})^{2}\overbar{\beta}\overbar{\delta_{SO_{4}^{-2}}}
\end{equation}
Now, $\overbar{\delta_{SO_{4}^{-2}}}$ can be obtained from the molar scattering cross section ($\alpha_{SO_{4}^{-2}}$), the mean column burden of the aerosol layer $\overbar{B_{SO_{4}^{-2}}}$ in $mol~m^{-2}$, and relative increase in scattering due to accretion of water vapor on the surface of aerosols represented as \textit{f}(RH) (where RH is relative humidity).   
\begin{equation}
\overbar{\delta_{SO_{4}^{-2}}}=\alpha_{SO_{4}^{-2}}f(RH)\overbar{B_{SO_{4}^{-2}}}
\end{equation}
The sulfate burden is related to the rates at which aerosol is introduced in the atmosphere and removed from it.
\begin{equation}
\overbar{B_{SO_{4}^{-2}}}= \frac{Q_{SO_{2}} Y_{SO_{4}^{-2}} \tau_{SO_{4}^{-2}} } {A}
\end{equation}
where $Q_{SO_{2}}$ (in grams of S per year) is the $SO_{2}$ introduced into atmosphere by anthropogenic and natural processes, $Y_{SO_{4}^{-2}}$ is the fractional $SO_{2}$ that is oxidized to produce ${SO_{4}^{-2}}$, $\tau_{SO_{4}^{-2}}$ is the lifetime of sulfate in the atmosphere, and A is the area of earth's surface. 

The values of these constants used in the scheme \cite{Charlson1992} have been listed in Table. \ref{tab:sul_dir_tab}. Some of the values used originally have been revised to reflect our current understanding. For example, the radiative efficiency for clear sky has been reduced to 25 $W/m^2$ from the originally used 83 $W/m^2~$ \cite{Stevens2014}. Further, the clear sky ratio has been increased to 0.6 from 0.4 as aerosols significantly scatter light in areas with sparse clouds.   

\begin{table}
\centering

\caption{Parameter values for direct radiative forcing of aerosols\cite{Charlson1992} $~$ \cite{Stevens2014}. }
\label{tab:sul_dir_tab}
\begin{tabular}{lllll}
 \toprule
 Quantity & Value & Units\\
 \midrule
 $Q_{SO_{2}}$  & Time-series Data \cite{Stern2005} & g of Sulfur per year \\
 $Y_{SO_{4}^{-2}}$ & 0.62(revised) & \\
 $\tau_{SO_{4}^{-2}}$ & 0.01~(revised) & year \\
 A & $5*10^{14}$ & $m^{2}$ \\
 $\alpha_{SO_{4}^{-2}}$ & 5 & $m^{2}~(g~of~{SO_{4}^{-2}})^{-1} $ \\
 \textit{f}(RH) & 1.7 & \\
 T & 0.38~(revised) &  \\
 $(1-A_{c})$ & 0.6~(revised) & \\
 $(1-\overbar{R_{s}})$ & 0.85 & \\
 $\beta$ & 0.29 & \\
 \bottomrule
\end{tabular}
\end{table}

In an attempt to test the impact that Galactic Cosmic Rays (GCRs) have on our climate, we have developed a toy model to vary 'k'. This is in deviation from prior models which have assumed it to be 1. The model to vary 'k' has been detailed in later sections.\\
\\
\textit{For indirect radiative forcing}

A 15\% increase in droplet number concentration causes a increase in global mean albedo ($R_{GM}$) of 0.003 \cite{Charlson1992}. But this has been revised to 0.001 because spatial heterogeneity reduces the effective cloud cover that interacts with aerosols \cite{Stevens2014}. The relation is described below. 
\begin{equation}
\overbar{\Delta \alpha}=k*A_{mst}*0.8*0.083*\log\frac{N}{N_{0}}
\end{equation}
where  $A_{mst}$ is the fraction of globe that is covered by marine stratiform clouds and N is the CDNC. Though $A_{mst}$ is around 0.3, its effective value is expected to be 0.1 \cite{Stevens2014}. 

The empirically derived relation between CDNC and sulfate concentrations \cite{Boucher1995} is
\begin{equation}\label{eq:sulf_indir_1}
CDNC = 10^{2.21+0.41\log(m_{SO_{4}^{-2}})}
\end{equation}
where CDNC is expressed in per cubic centimeter and $SO_{4}^{-2}$ levels in micro-grams per cubic meter. As the time series data \cite{Stern2005} is expressed in Gg (giga-grams) of sulfur per year, the following scaling can be used to convert the units. 

\begin{equation}\label{eq:sulf_indir_2}
m_{SO_{4}^{-2}}= \frac{emission~rate*\tau*Y_{SO_{4}^{-2}}*3*10^{6}}{365*Volume~of~atmosphere} 
\end{equation}
\\where volume of atmosphere is taken to be $5.1*10^{18}~m^{3}$ and the factor of 3 is due to the mass-ratio of sulfate to sulfur.

The change in albedo thus obtained using the above two formulations influence the TSI entering the climate system. The albedo is included in the model using eq. \ref{eq:Inet}.\\
\\
\textit{Stratospheric aerosol from volcanic explosions}

\begin{figure}[h]
\captionsetup{singlelinecheck = false, justification=justified}
\includegraphics[width=\linewidth]{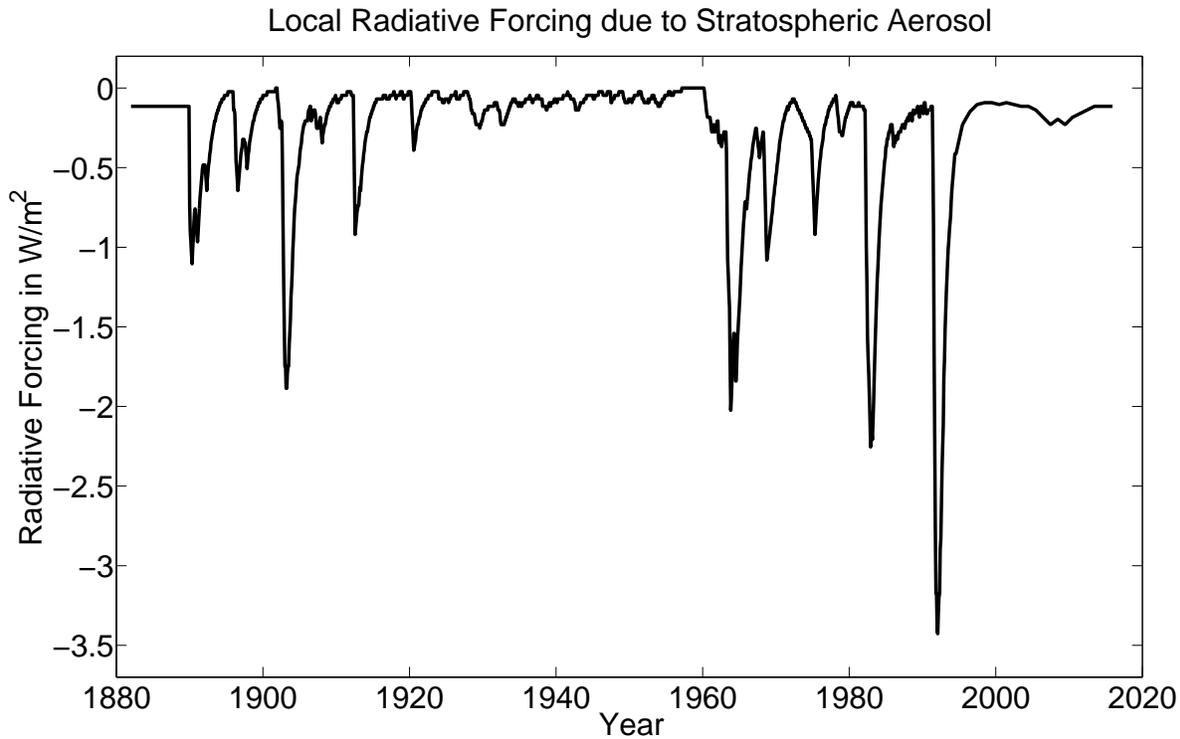}
\caption{Volcanoes are sometimes cataclysmic enough to introduce large quantities of sulfur into the stratosphere. The aerosols thus introduced lead to optical thickness in the shortwave spectrum.}
\label{fig:stratospheric_aerosols}
\end{figure}

Apart from the troposphere burdening of sulfur, volcanoes have also been attributed to sulfur burdening in the stratosphere. In the absence of clouds, the main parameter causing radiative forcing is the aerosol optical depth \cite{Lacis1992}. Volcanic aerosol can be significantly stronger than greenhouse forcing (in their local environment) and is an important part of our model. Various indices have been used to obtain an aerosol optical depth time series from 1850-1990 \cite{Sato1993}. We have used an updated time series in our model which includes satellite observations. The authors estimated the radiative forcing due to optical depth to be
\begin{equation}
\Delta F_{R} \sim - 30 \tau
\end{equation}
where $\tau$ refers to the optical depth at 500nm. This was later revised by \cite{Hansen2005} to 
\begin{equation}
\Delta F_{R} \sim - 23 \tau
\end{equation}
The relation is expected to be non-linear for large eruptions such as that of Mt. Pinatubo. However, we have assumed a linear relation as it holds for most of the time series. The local radiative forcing has been illustrated in Fig. \ref{fig:stratospheric_aerosols}.

This correction is introduced by subtracting from the incident radiative forcing [I(t)] that enters the climate system.

\subsubsection*{Greenhouse Gasses}
As opposed to the aerosol forcing which cools the earth's atmosphere, greenhouse forcing acts along the third pathway and increases its heat content \cite{houghton1992}. This process remains one of the most widely studied aspects of our climate system and has been attributed to causing the global warming effect. The increased forcing along this pathway is due to the increase in greenhouse gas concentrations in the troposphere. 

As the greenhouse gasses are rather well understood, we have directly used the radiative forcing due to them in our calculations [G(t)]. The forcing terms, functional forms for said forcing and coefficients have been listed in Table.\ref{tab:ghg_forcing}   $~$\cite{ipcc1990}$~$\cite{Hansen1988}. We have used the levels for $CO_{2}$, $CH_{4}$, $N_{2}O$, CFC-11 and CFC-12 for our calculations as they contribute over 99\% of the total forcing.  

\begin{figure}
\captionsetup{singlelinecheck = false, justification=justified}
\includegraphics[width=\linewidth]{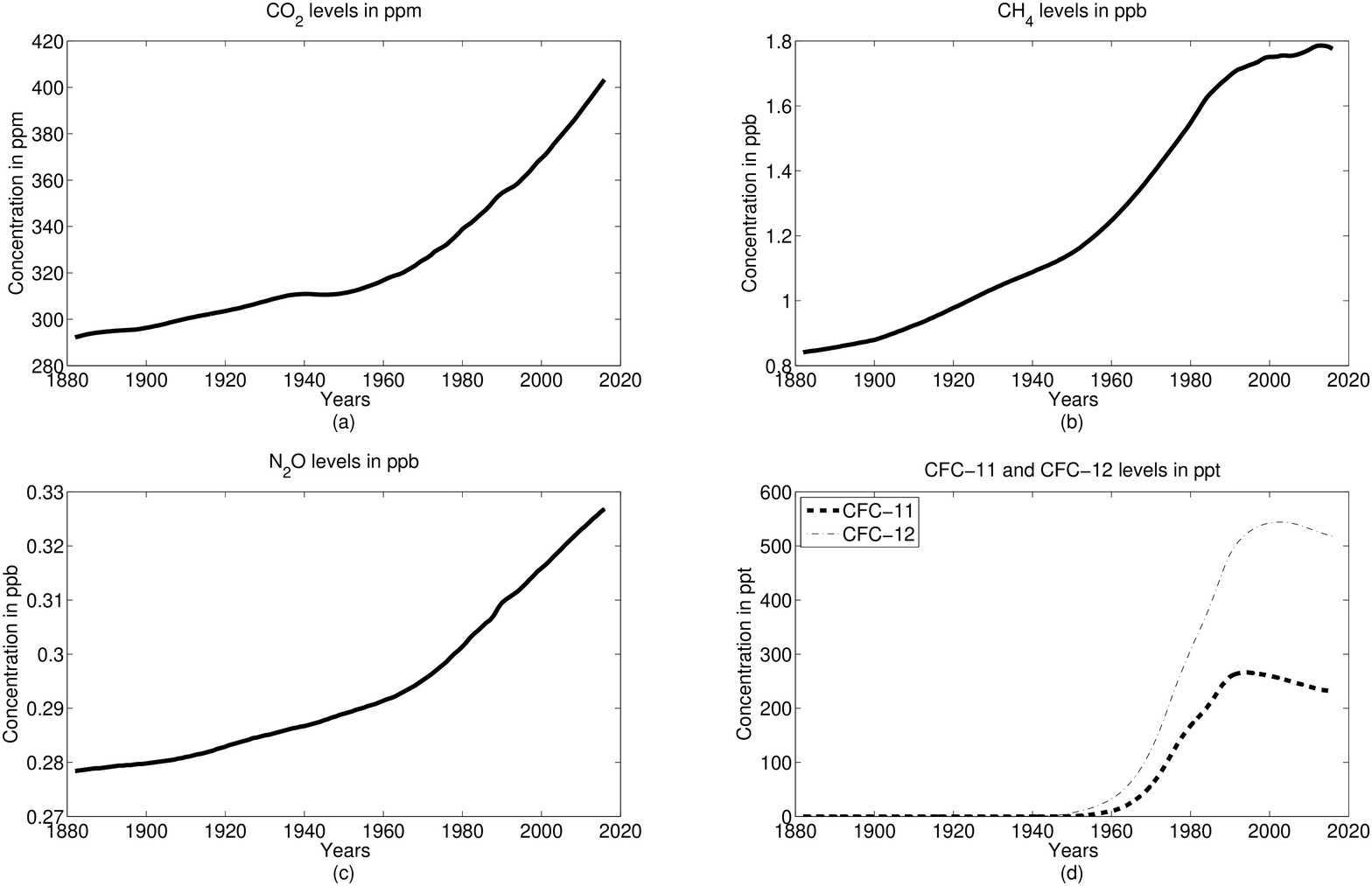}
\caption{Greenhouse Gas levels for 1882-2015. The time-series has been obtained from RCP 2.6 non interacting gas levels.}
\label{fig:ghg}
\end{figure}

\begin{table}[h]
\centering
\caption{Expressions for radiative forcing due to greenhouse gases}
\label{tab:ghg_forcing}
\begin{tabular}{lllll}
 \toprule
 Trace Gases & Expression for Radiative forcing & Description\\
 \midrule
 $CO_{2}$ & $\Delta F=6.3ln(C/C_{0})$ & where C denotes the \\
 & & concentration in ppm.\\
 & & Valid for C\textless 1000 ppm.  \\
 $CH_{4}$ & $\Delta F=0.036(\sqrt{M}-\sqrt{M_{0}})$  & where M is $CH_{4}$ \\
 &$~~~~~~~~~-(f(M,N_{0})-f(M_{0},N_{0}))$ & concentration in ppb.\\
 & & M\textless5ppb. \\
 $N_{2}O$ & $\Delta F=0.14(\sqrt{N}-\sqrt{N_{0}})$  & where N is $N_{2}O$ \\
 &$~~~~~~~~~-(f(M_{0},N)-f(M_{0},N_{0}))$ & concentration in ppb.\\
 & & N\textless5ppb. \\
 CFC-11 & $\Delta F= 0.22(X-X_{0})$ & where X is expressed \\
 & & in ppb.\\
 & & $(X-X_{0})$\textless 2ppb. \\
 CFC-12 & $\Delta F= 0.28(Y-Y_{0})$ & where Y is expressed \\
 & & in ppb.\\
 & & $(Y-Y_{0})$ \textless 2ppb.\\ 
 \bottomrule
\end{tabular}\\
here $f(M,N)=0.47ln(1 + 2.01*10^{-5}(MN)^{0.75} + 5.31*10^{-15}M(MN)^{1.52} )$
\end{table}
The time-series data for $CO_{2}$, $CH_{4}$ and $N_{2}O$ used were Representative Concentration Pathways (RCP) 2.6 non-interacting gas levels \cite{nazarenko2015} and can be accessed from the GISS website (Goddard Institute of Space Studies). CFC-11 and CFC-12 data were obtained separately \cite{Bullister2015}. The greenhouse gas levels have been mapped in fig.\ref{fig:ghg}.

\subsubsection*{Galactic Cosmic Rays}\label{ssec:GCR}
GCRs are speculated to impact the climate by modulating the formation of clouds. Different mechanisms of GCR interactions with the atmosphere have been proposed. According to one such mechanism, it is expected that highly energetic GCRs cause condensation by a process known as electrofreezing  \cite{Pruppacher1973}$~$\cite{Abbas1969}. The occurrence of such a process in the atmosphere has been highly debated. Another process that is speculated to occur during the passage of GCRs through our atmosphere is the increase in nucleation rate of $SO_4^{-2}~$ \cite{Kirkby2011}. We test the later idea by introducing a simple mathematical formulation that encodes the effect of ion-induced nucleation due to GCRs.

It has been experimentally demonstrated that ion-induced nucleation of sulfuric acid increases at least twofold (can range up to tenfolds or more) for ground-level GCR as compared to neutral nucleation \cite{Kirkby2011}. We postulate that the nucleation rate increases at minimum solar activity and drops at maximum solar activity due to a corresponding increase and decrease in GCR. This is because the GCRs with the energy levels that lead to increase in nucleation are expected to be trapped by the solar wind during high solar activity \cite{balasubrahmanyan1969}. This results in
\begin{equation}\label{eq:toy}
k(t)= 0.2+\left( 0.8*Fractional~Increased~Intensity~of~GCRs \right)
\end{equation}	
We then use the inverse relation between GCRs and solar activity to define the following equation
\begin{equation}
Fractional~Increased~Intensity~of~GCRs = \frac{I_{max}-I(t)}{I_{max}-I_{min}}
\end{equation}
where $I_{max}$ and $I_{min}$ refers to the maximum and minimum TSI. We have used the value of TSI as a proxy for solar activity.

\subsubsection*{Total Solar Irradiance}
Total Solar Irradiance [I(t)] is the power per unit area incident upon the top of earth's atmosphere due to the sun's luminous output. The energy is received in the form of electromagnetic radiation and is measured across all wavelengths. 

The composite total solar irradiance values are obtained from NOAA Climate Data Record \cite{Coddington2015} and have been used as a source term in our model.

\subsubsection*{El-Nino Southern Oscillations}
Several indices have been traditionally used to study the El-Nino Southern Oscillations (ENSO) \cite{hanley2003}. Of these, the multivariate ENSO index uses six different variables as an indicator for studying the oscillations. However, as we are only interested in using energy input/output in our model, we have used the temperature anomaly in the Nino 3.4 region \cite{Rayner2003} as a representative of the ENSO phenomenon. We have then calculated the energy imbalance created by the regions which contribute to the ENSO using basic physical formulas of radiative loss. We express this energy imbalance as:

\begin{equation}\label{eq:ENSO1}
E(t)=\epsilon*4*\sigma*\frac{A}{16}*T_e^3*dT_a
\end{equation}
where $T_e$ is the mean temperature in the Nino 3.4 region for data obtained from 1951-2000 \cite{Rayner2003}. $dT_{a}$ is the temperature anomaly mentioned above. We have further approximated the area of the earth's surface that contributes to this process is 1/16th of its total area. This is estimated from the fact that the area of the earth where this phenomenon occurs is approximately spread across 120 degrees of longitude and 10 degrees of latitude on each side of the equator. $\epsilon$ is the fraction of infrared radiation absorbed by the atmosphere.

However, the El-Nino period is usually accompanied by thick clouds over the Pacific Ocean. Therefore, we expect that all the extra energy in the form of infrared radiation emitted by the ocean surface (in the ENSO region) during this period will be absorbed by the atmosphere. During the El-Nino phase, we have assumed $\epsilon \approx 1$ \cite{radel2016}. However, the actual fraction of infrared radiation absorbed will be above the global mean but below 1. 

The time delay due to ENSO cannot be captured spontaneously in our model as the time period related to ENSO oscillations varies erratically between 2 to 7 years \cite{kestin1998}. After the correction for ENSO is included, the delay due to the remaining oceanic circulation belts can be obtained.

\bibliography{references}

\section*{Acknowledgements}
This model was conceived and developed by CESSI, a multi-institutional Center of Excellence established and funded by the Ministry of Human Resource Development, Government of India.The authors would like to thank Dr. Judith Lean for sharing forcing data time series that made this study possible.

\section*{Author contributions statement}
R.T. performed the study and suggested the incorporation of GCR modulation. D.N. suggested the idea of using delay equations to study the climate and guided the study. All authors reviewed the manuscript. 

%\section*{Additional information}

%\textbf{Accession codes}: The authors do not declare any accession codes.\\
%\textbf{Competing financial interests}: The authors declare no competing financial interests. 

\end{document}